\documentclass[conference]{IEEEtran}
\IEEEoverridecommandlockouts
\usepackage{cite}
\usepackage{amsmath,amssymb,amsfonts}
\usepackage{algorithmic}
\usepackage{graphicx}
\usepackage{textcomp}
\usepackage{xcolor}
\usepackage[T1]{fontenc} 
\def\BibTeX{{\rm B\kern-.05em{\sc i\kern-.025em b}\kern-.08em
    T\kern-.1667em\lower.7ex\hbox{E}\kern-.125emX}}
\usepackage{fancyhdr}

\begin{document}

\title{Multivalued circuits and Interconnect issues }

\author{\IEEEauthorblockN{ Daniel Etiemble}
\IEEEauthorblockA{\textit{Computer Science Laboratory (LRI)} \\
\textit{Paris Saclay University}\\
Orsay, France \\
de@lri.fr}

}
\maketitle

\begin{abstract}
Many papers have presented multi-valued circuits in various technologies
as a solution to reduce or solve interconnection issues in binary circuits.
This assumption is discussed. While 4-valued signaling
could divide by two the number of interconnects between building blocks, it turns out
that circuit designers use interconnect standards based on differential pairs such as PCIe,
Infiniband, RapidIO, NVLink, etc. Doubling the number of binary signals is  a better
solution than using single-ended quaternary signals.
The design of quaternary basic gates, adders and multipliers are compared with the corresponding binary ones.
 At each level, the transistor count ratio between quaternary and binary circuits is  greater
than the x2 information ratio between base 4  and base 2.
Quaternary signaling is not a solution, either between  or within  circuit blocks . 
\end{abstract}


\section{Introduction}\label{sec1}
\begin{figure}[htbp]
\centerline{\includegraphics  [width =7 cm]{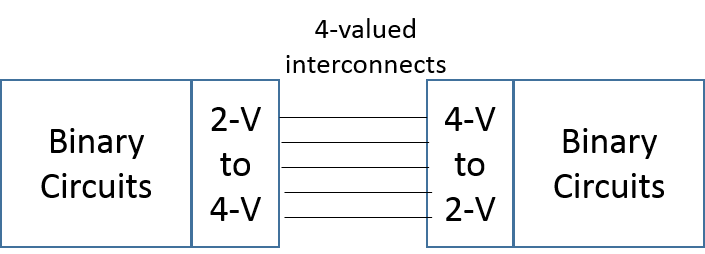}}
\caption{4-valued interconnects between binary circuits}
\label{4Vint}
\end{figure}

\begin{figure}[htbp]
\centerline{\includegraphics  [width =4 cm]{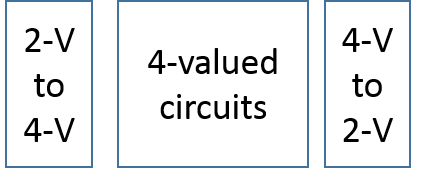}}
\caption{4-valued  circuits in a binary environment}
\label{4VC}
\end{figure}

Many published papers proposing new m-valued circuits justified their proposal by interconnection issues in binary circuits. Just to present one among a lot of quotes, "One of the main problems in binary logic is the high volume of interconnections which can increase the chip area and power consumption" \cite{Sha}. It is quite evident that m-valued signals carry more information than binary signals and that less m-valued signals would be needed than binary ones to carry the same amount of information. However, if the objective is to reduce the interconnect issues of binary circuits, 4-valued circuits are the best solution as they allow a simple interface between binary and 4-valued circuits.
There are two different options:
\begin{itemize}
\item Using binary circuits and using 4-valued signaling between binary circuits, as shown in Figure \ref{4Vint}. In that case, the number of interconnects between the binary parts is divided by 2.
\item Using 4-valued circuits in a binary environment, as shown in Figure \ref{4VC}. In that case, the 4-valued circuits are based on 4-valued operators with  4-valued interconnects.

\end{itemize}


\section{Methodology}
In the last decade, most proposed designs of m-valued circuits use simulations of CNTFET technologies. The CNTFET circuitry is similar to the CMOS one while presenting a significant advantage: the threshold voltage varies as the inverse function of the diameter of the tube. Different diameters provide the different threshold voltages that are needed in m-valued circuits.

 However, this technology  is far from being mature. In 2019, a 16-bit RISC microprocessor has been built with 14,000 CNFET transistors \cite{Hills}. While this is a significant advance for CNTFET technology, we may observe that the Intel 8086 CPU, which was a 16-bit microprocessor, has been launched in 1978 with 29,000 transistors, more than 40 years ago! As of 2019, the largest  transistor count in a commercially available microprocessor was 39.54 billion MOSFETs, in AMD's Zen 2 fabricated using TSMC's 7 nm FinFET semiconductor manufacturing process. As of 2020, the largest transistor count in a GPU (NVidia Ampere) was 54 billion transistors with the same 7 nm process. The CNTFET 16-bit microprocessor manufacturing process had 5 metal layers, while the number of metal layers in nano-CMOS technologies ranges from 8 to 15, with a trade-off between integration and cost. So, comparing m-valued circuits using CNTFET technologies with the most recent FinFET or FDSOI  technologies would be a nonsense.

\begin{figure}[htbp]
\centerline{\includegraphics  [width =8 cm]{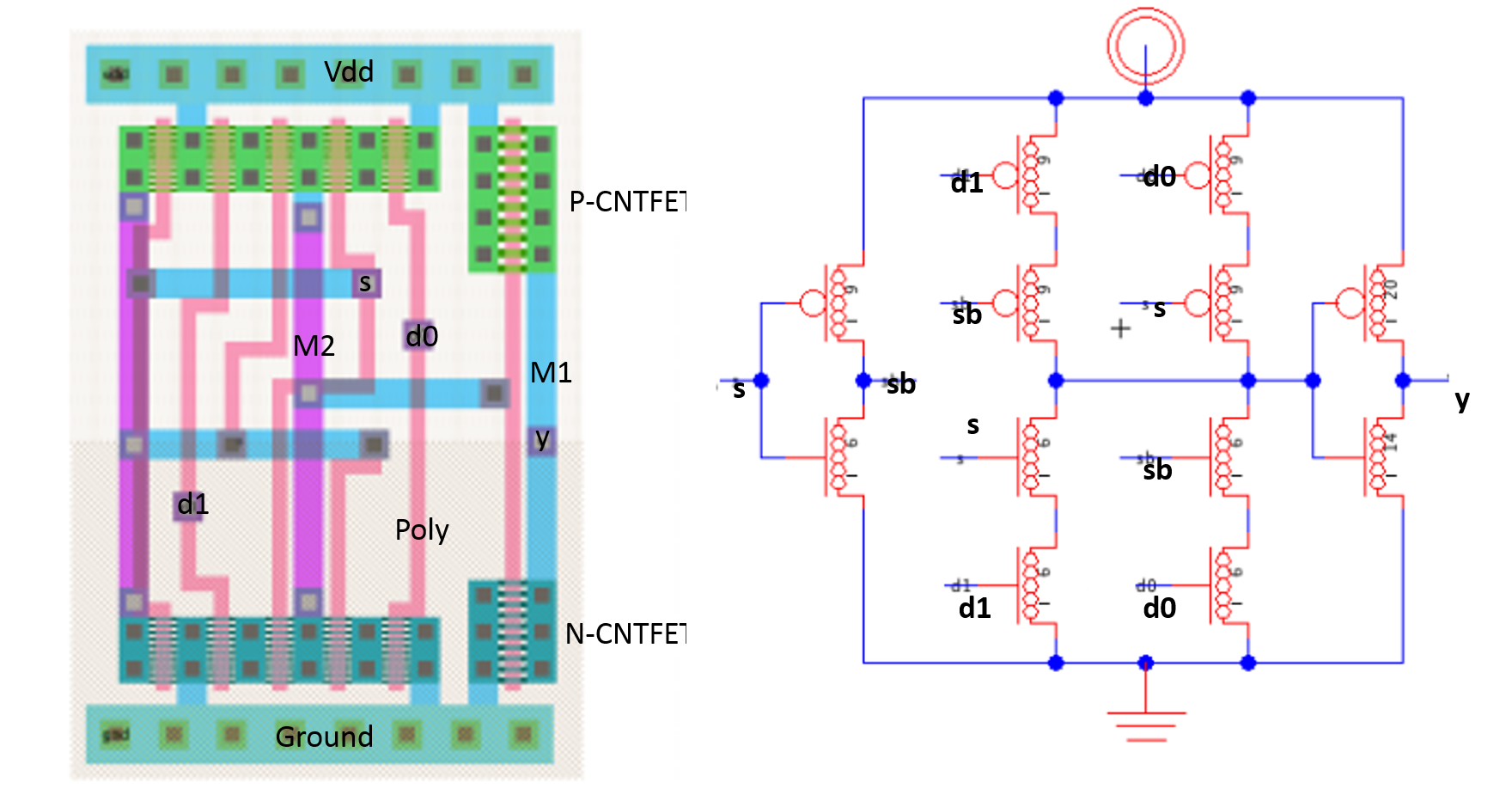}}
\caption{2:1 CNTFET multiplexer. Layout (left) and circuit (right)}
\label{Layout}
\end{figure}

As the point is to compare the interconnections between binary and 4-valued circuits, we use the CNTFET technology as a basis for all the comparisons. Directly comparing interconnects is impossible. While most 4-valued proposals present the electrical scheme, they generally neither provide any layout information nor even the electrical scheme of the corresponding binary circuits. While interconnects are represented on a plane in the electrical scheme, they are implemented according to several layers in the IC layout.  Figure \ref{Layout} presents the layout and the circuit scheme of a 2-input multiplexer \cite{Huang}. All the connections are not shown in the electrical scheme. The connections are implemented as Metal 2 (blue), Metal 1 (violet) and polysilicon (pink).  However, considering the connections at the transistor level gives insight on the interconnection at the layout levels. There are few possibilities that more connections at the transistor levels lead to less interconnections and less chip area at the layout level. If there are significant differences between number of connections at transistor level between m-valued and binary circuits, we can easily assume that the situation is similar at the layout level. It is also very doubtful that many more transistors could lead to less interconnects. The ratio m-valued transistor count/binary transistor count therefore gives some indication.

\section{4-valued interconnects between binary circuits}
\subsection{Telecom applications}
\begin{figure}[htbp]
\centerline{\includegraphics  [width = 8 cm]{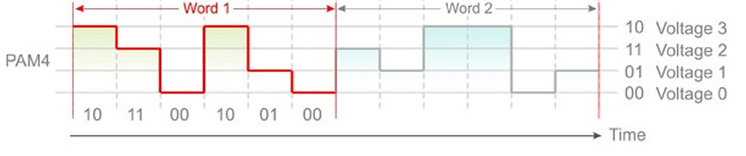}}
\caption{PAM4  encoding}
\label{PAM4}
\end{figure}

There is one significant application of this approach. Reducing the number of
interconnects with multiple levels is used in amplitude
modulation: for instance, PAM-4 coding \cite{PAM4} (Figure \ref{PAM4}) that uses 4
levels to code 2 bits is adopted for high-speed data
transmission (IEEE802.3bs).

However, telecom industry and computer industry are quite different.
\subsection{Digital electronics and computer industry}
\subsubsection{Dividing by 2 the number of interconnects}
Different papers have been published in which 4-valued signaling is used
between binary circuits in different technologies: TTL \cite{etiTTL}, ECL  \cite
{etiECL}, CMOS, CNTFET, etc.  2-valued to 4-valued encoder and 4-valued to 2-valued
decoder circuits must be designed for each technology. 

\begin{figure}[htbp]
\centerline{\includegraphics  [width =8 cm]{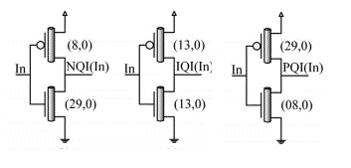}}
\caption{Threshold detector circuits presented in \cite{Ebrahimi}}
\label{Tdetectors}
\end{figure}

\begin{figure}[htbp]
\centerline{\includegraphics  [width =6 cm]{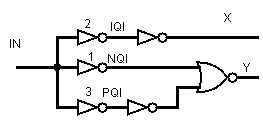}}
\caption{Quaternary to binary decoder (Gray code)}
\label{QBGray}
\end{figure}

\begin{figure}[htbp]
\centerline{\includegraphics  [width =4 cm]{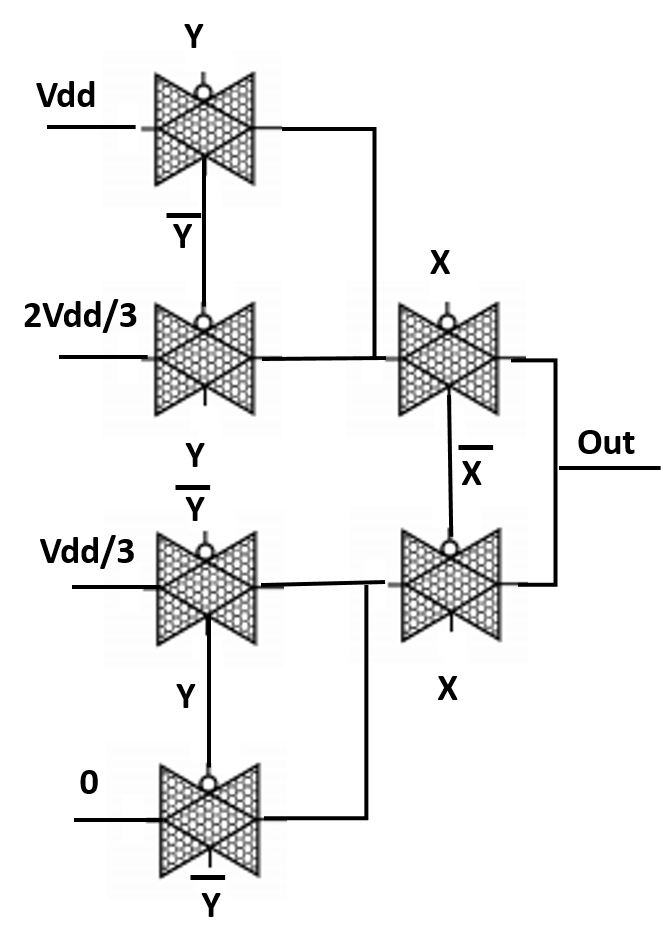}}
\caption{Binary to quaternary encoder (Gray code)}
\label{BQGray}
\end{figure}

With CNTFET technology, a possible implementation is now presented. The truth table of the decoder using Gray code is presented in Table \ref{NQI}. The threshold detector circuits implementing functions NQI, IQI and PQI are shown in Figure \ref{Tdetectors}. The corresponding 4 to 2 decoder with 14 T is presented in Figure \ref {QBGray}. A possible implementation of the 2 to 4 encoder is presented in Figure \ref{BQGray}. It has 12 T.

\begin{table}
\centering
\caption{Gray 4 to 2 decoder}
\begin{tabular}{|c||c|c|c||c|c||}
  \hline
 Q&NQI&IQI&PQI&X&Y\\
\hline
 0&3&3&3&1&0\\
 1&0&3&3&1&1\\
2&0&0&3&0&1\\
3&0&0&0&0&0\\
  \hline
\end{tabular}
\label {NQI}
\end{table}

\subsubsection{Doubling the number of interconnects}
\begin{figure}[htbp]
\centerline{\includegraphics  [width = 7 cm]{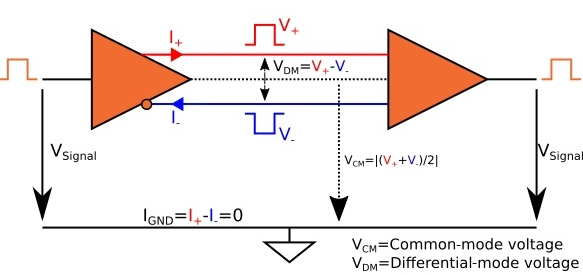}}
\caption{Differential signaling}
\label{Dif}
\end{figure}

 It turns out that 4-valued interconnects have not been used by IC designers.
For "long distance" or "high-speed" transmissions, the solution has been to use differential signals instead of single end signals, as shown in Fig. \ref{Dif}. Differential signaling has advantages over single-ended such as:
\begin{itemize}
\item Better noise margins, as the signal swing is the difference between positive and negative signals
\item Self-reference as the threshold is $(positive signal + negative signal)/2$
\item Reduced switching time due to lower voltage swing.
\item Subtraction between positive and negative signals rejects common-mode noise.
\end{itemize} 

While differential links are used at physical level, different buses and links have been defined and used in computer architectures. There are many examples: SuperSpeed USB  3.0 \cite{USB}, PCIe \cite{PCI}, XAUI \cite{XAUI}], InfiniBand \cite{InfiniBand}, RapidIO \cite{RapidIO} , and SATA  \cite{SATA}. We only give some more details for two examples:
\begin{itemize}
\item PCI Express: It is a high-speed serial bus. A PCI express link (or interconnect) between two devices consists in a number of lanes, each lane being composed of two differential signaling pairs, one for receiving data and the other one for transmitting. Successive versions have been defined. PCI Express 1.0 has been introduced in 2003 and PCI Express 5.0 is expected in 2022.
\item NVLink: To improve the interconnection bandwidth between CPU and GPUs, NVidia introduced a new interconnect architecture called NVLink \cite{NVlink}. A single NVLink is a bidirectional interface incorporating 32 wires forming eight differential pairs in each direction. The first implementation NVLink1 was introduced in the Tesla P100 GPU (2016). Each link has a 40 GB/s bidirectional bandwidth. NVLink1 is also used by IBM in the Power8 microarchitecture. NVlink2 was introduced with GV100 GPU (2018). Bidirectional bandwidth is now 50 GB/s. 
\end{itemize}

\section{4-valued interconnects within 4-valued circuits}
In this section, we examine whether 4-valued circuits reduce the number of interconnects compared to the corresponding binary circuits. According to Shannon theory of information, when N events have the same probability to occur, the corresponding amount of information is $ I=log_2(N)$ bits (or shannons). When N=2, I=1 bit and when N=4, I=2 bits. A quaternary wire carries two times the amount of information of a binary one. It means that the information ratio is 2. A quaternary circuit will be more efficient than a binary one if its complexity is not greater than 2. 
Comparing quaternary circuits and the corresponding binary ones is not easy. Which technology do we consider? How to evaluate complexity?

\subsection{Quaternary logic gates}
\begin{figure}[htbp]
\centerline{\includegraphics  [width = 5 cm]{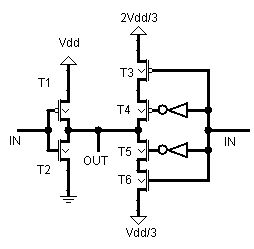}}
\caption{A 4-valued CNTFET inverter}
\label{4VINV}
\end{figure}

Inverters are the first gate to consider. Figure \ref{4VINV} shows a 4-valued inverter presented in \cite {Sharifi}. This scheme allows a quick comparison with binary circuits. A 4-valued inverter carries the same amount of information than two binary inverters: 2 bits. Right part of Figure 2 includes two binary inverters, which means that the six supplementary transistors are just the overhead of the 4-valued inverter compared to two binary inverters. All the connections between these transistors are overhead, which doesn't exist for binary implementation. It should also be noticed that the 4-valued implementation uses three power supply rails ($V_{dd}, 2V_{dd}/3, V_{dd}/3$) instead of one. Power supply distribution at the layout level is much more complicated. The quaternary inverter presented in \cite{Ebrahimi} has also 10 transistors.

Obviously, the situation is similar for NOR, NAND and more complex gates. Table \ref{QNAND-TC} compare the transistor count of two different 4-valued NAND with the  equivalent binary NAND. For processing 2 bits of information, the quaternary Nands have at least two times more transistors. The results are the same for NOR gates.
\begin{table}
\centering
\caption{2-input Nand gate transistor count}
\label{QNAND-TC}
\begin{tabular}{|c|c|c|c|c|c|}
\hline
2-input Nand  & 4-V Nand \cite{Sharifi}& 4-V Nand\cite{Ebrahimi}&Binary Nand  \\
\hline
Transistor count  &20&16&4 \\
\hline
4V/2V TC ratio &5&4&1\\
\hline
\end{tabular}
\end{table}
\subsection{Quaternary full adders} \label{ENCDEC}
A detailed examination of the implementation of quaternary adders has been presented in \cite{etiQADD}. They are based on the following versions:
\begin{itemize}
\item V1: Interfacing a 2-bit binary adder with a 4-to-2 decoder and a 2-to-4 encoder.
\item V2: Direct implementation of Table \ref{T1}.  A function f(inputs) is decompose into f(inputs) = 3.f3 + 2.f2 + 1.f1 where f3, f2 and f1 are respectively the binary functions of the inputs for which the functions have values 3, 2 and 1. f3, f2 and f1 includes the NQI, IQI and PQI functions of input variables (Table \ref{NQI}). This approach is used in the adder presented in \cite{Ebrahimi}.
\item V3:  This intermediate approach uses multiplexers to implement subfunctions that can be derived from Table \ref{T1}. An example of subfunction is the successor function: When A = 1 and Ci = 0 then QS = (B+1) mod. 4. This approach is used in adders designed in \cite {Moaiyeri} and \cite {Roosta}. Results  \cite {Roosta} are presented with 1 and 3 power supplies. 
\end{itemize}

\begin{table}
\centering
\caption{Truth table of a quaternary adder}
\begin{tabular}{|c|c|c||c|c|c|c|c|c||c|c|}
  \hline
A&B&Ci&QS&QC& &A&B&Ci&QS&QC\\
\hline
 0&0&0&0&0&&0&0&1&1&0\\
0&1&0&1&0&&0&1&1&2&0\\
0&2&0&2&0&&0&2&1&3&0\\
0&3&0&3&0&&0&3&1&0&1\\

 1&0&0&1&0&& 1&0&1&2&0\\
1&1&0&2&0&&1&1&1&3&0\\
1&2&0&3&0&&1&2&1&0&1\\
1&3&0&0&1&&1&3&1&1&1\\

 2&0&0&2&0&& 2&0&1&3&0\\
2&1&0&3&0&&2&1&1&0&1\\
2&2&0&0&1&&2&2&1&1&1\\
2&3&0&1&1&&2&3&0&2&1\\

 3&0&0&3&0&&3&0&1&0&1\\
3&1&0&0&1&&3&1&1&1&1\\
3&2&0&1&1&&3&2&1&2&1\\
3&3&0&3&1&&3&3&1&3&1\\
  \hline
\end{tabular}
\label {T1}
\end{table}

\begin{table}
\centering
\caption{NQ1, IQI and PQI funtions}
\begin{tabular}{|c||c|c|c||c|c|}
  \hline
 Q&NQI&IQI&PQI\\
\hline
 0&3&3&3\\
 1&0&3&3\\
2&0&0&3\\
3&0&0&0\\
  \hline
\end{tabular}
\label {NQI}
\end{table}

\begin{figure}[htbp]
\centerline{\includegraphics  [width = 7 cm]{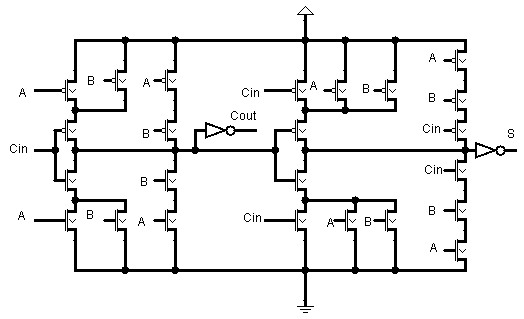}}
\caption{28T binary full adder}
\label{28T}
\end{figure}

For a fair comparison, we assume that the binary full adder uses the conventional 28 T design (Figure \ref{28T}). The binary XOR gates used in V1 have 10T, which is typical of standard cell library implementation.

Table \ref{Adders} present the transistor count of the three different 4-valued full adders and the transistor count ratio with the conventional binary full adder. The TC ratio is always greater than 2, which is the information ratio. Except the quaternary adder  presented \ in \cite{Roosta} with 3 power supplies, the TC ratio is greater or equal to 4. The direct interfacing of 2-bit binary adder with 4 to 2 decoders and 2 to 4 encoder is close to the best quaternary versions. We outline that the results for version 1 are somewhat pessimistic (conventional implementation of XOR and 1-bit full adder) while version 3  \cite{Roosta} are somewhat optimistic. As detailed in \cite{etiQADD},when reducing the high fan out supported by some inverters, the TC raises to 100 (1 power supply) and 148 (3 power supplies).
 
\begin{table}
\centering
\caption{Adder transistor count and 4V/2V transistor ratios}
\label{Adders}
\begin{tabular}{|c|c|c|c|c|c|c|}
\hline
4-V adders&V1&V1& V2\cite{Ebrahimi}&V3\cite{Moaiyeri} &V3\cite{Roosta}&V3\cite{Roosta} \\
&3  PS&1 PS&&&3  PS&1 PS\\
\hline
TC &112 &112&111&154&82&130\\
\hline
TC ratio&4 &4&4&5.5&2.9&4.65\\
\hline
\end{tabular}
\end{table}

N-digit quaternary ripple carry adders are implementing by concatenating 1-digit quaternary adders. A N-digit quaternary adder must be compared with a 2N-bit binary adder. As the TC ratio is always greater than 2, the quaternary adders always use  two times more transistors for all published quaternary adders (except \cite{Roosta}) than the corresponding binary ones. More transistors mean more interconnects. Version 1 is a good example: the decoder and encoder circuits are just overhead over the binary implementation.
As shown in \cite{etiQADD}, quaternary carry computation of a 4-digit adder is slightly more efficient than binary carry computation of 8-bit adder, either for Carry-Look Ahead Adders or Carry-Save Adder. However the gain is too small to compensate  the transistor count difference between quaternary and binary full adders.

\subsection{Quaternary multipliers}
\begin{figure}[htbp]
\centerline{\includegraphics  [width =8 cm]{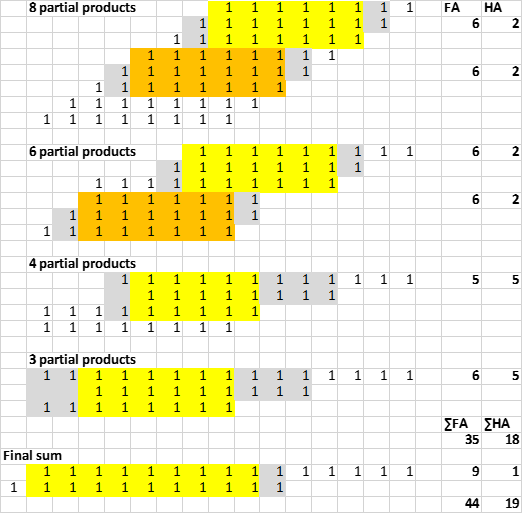}}
\caption{8*8 bit multiplication with a Wallace tree}
\label{88WT}
\end{figure}

\begin{figure}[htbp]
\centerline{\includegraphics  [width =8 cm]{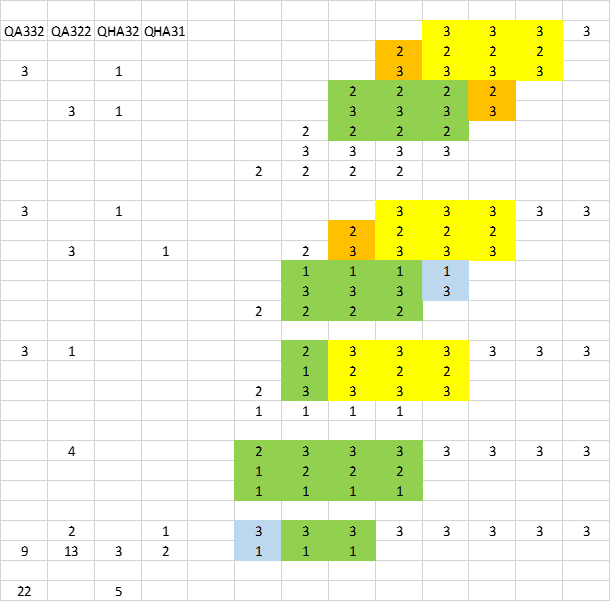}}
\caption{4x4 quaternary multiplier with  a Wallace Tree}
\label{WQ44}
\end{figure}

MVL designers generally only propose the implementation of a 1-digit quaternary multiplier. Sometimes, N*N quaternary multipliers are proposed \cite{Hajare}. In both cases, no comparison is done with the corresponding binary multipliers.
As detailed in \cite{etiQMUL}, the best implementation of a quaternary 1-digit multiplier uses the technique quoted in version 3.
The corresponding transistor count ranges from 54 T up to 76 T. It depends on the fan-out supported by some inverters. The corresponding 1-bit multiplier is implemented by an AND gate (6 T). However, a direct comparison cannot be done as an NxN combinational multiplier uses both 1-digit multipliers and 1-digit adders.
The simplest technique to implement NxN multipliers use the Wallace technique. For instance, Figure \ref{88WT} shows the Wallace reduction tree for a 8x8 bit multiplier. The situation is more complicated for the reduction of a 4x4 quaternary multiplier, as shown in Figure \ref{WQ44}:
\begin{itemize}
\item There are 16 1-digit multipliers versus 64 1-bit multipiers. However, the 1-digit multiplier is far more complicated than a simple AND gate.
\item Each line to be reduced has only 4 partial products instead of 8
\item However, a 1-digit multiplier delivers a quaternary product term and a ternary product term. With base 4, 3x3= 21, which means that the carry values are 0, 1 or 2. In Figure \ref{WQ44}, 3/2/1 corresponds respectively to quaternary/ternary/binary values. It means that different types of quaternary adders and half adders must be used in the reduction tree. They are named $Q_{332}$, $Q_{322}$,$ QHA_{32}$ and $QHA_{31}$ in Figure \ref{WQ44}. The corresponding designs are detailed  in \cite{etiQMUL}.
\end{itemize}
. 

A 4x4 quaternary multiplier may also be designed by interfacing a 8x8 binary multiplier with the 4 to 2 decoder and 2 to 4 encoder circuits used for the V1 quaternary adder (subsection \ref{ENCDEC}).
With conventional binary circuits and the optimistic version of \cite{Roosta}, the transistor count comparison for 8x8 binary multipliers and two 4x4 quaternary multipliers is given in Table \ref{Multipliers} 

\begin{table}
\centering
\caption{Quaternary multiplier transistor counts}
\label{Multipliers}
\begin{tabular}{|c|c|c|c|c|c|c|}
\hline
8x8 bit multiplier&4x4 Q multiplier&4x4 Q multiplier  \\&encoder and decoder&direct implementation\\\hline
1892 &2032 &2888\\
\hline
\end{tabular}
\end{table}

Significant conclusions can be derived from Table \ref{Multipliers}:
\begin{itemize}
\item Interfacing a binary multiplier with 4 to 2 decoder and 2 to 4 encoder circuits provides a better quaternary multiplier than the best direct implementation of a quaternary multiplier. 
\item Compared to the binary implementation of the multiplier, the encoder and decoder circuits are just an overhead that is needed to implement the best multiplier. It means the best quaternary multiplier has more interconnects that the corresponding binary one.
\end{itemize}

\subsection{Conclusion for logic gates and arithmetic circuits}
For basic logic gates, adders and multipliers, the best quaternary circuits have always more than 2 times the transistor count of the corresponding binary ones. If a quaternary wire carries two times the information of a binary one, it doesn't mean that quaternary circuits reduce the number of interconnects. The quaternary circuits are too much more complicated and actually augment the number of interconnects.

\section{Concluding remarks}
For more than 50 years, multi-valued circuits have been proposed as a potential solution to overcome interconnection issues. 4-valued circuits could divide by two the number of interconnects. It turns out that this assumption has been proved being false. Ironically, doubling the number of interconnects by using differential signaling pairs has been the solution that has been largely used to overcome the most significant interconnect issues. 
The lack of success of multivalued interconnects results from misunderstandings about both binary and multivalued circuits:
\begin{itemize}
\item The main issue of binary integrated circuits is not interconnections. It is power dissipation. While "Heat/Power wall" or "Memory wall" are well-known concepts for IC designers, "Interconnection wall" is not quoted. Increasing the number of metal layers in modern FinFET technologies reduces the interconnection issues. Reducing power consumption means using as small power supplies as possible, in the 0.8 V to 1V range. Within this range, there is no room for more than two different voltage levels. 
\item As shown in this paper for quaternary circuits, m-valued circuits are far more complicated than binary circuits in terms of transistor count and thus interconnects. The increased complexity is always greater than the increased computational power, expressed as $ log_2 (m)/log_2 (2)$. Multipliers are the most spectacular example.
\end{itemize}

\end{document}